\documentclass[aps,preprintnumbers,amsfonts,prb,twocolumn]{revtex4}

\renewcommand{\bbox}[1]{\mathbf{#1}}
\usepackage{psfig}
\usepackage[dvips]{graphicx}
\usepackage{psfrag}
\begin{document}
\newcommand{\z}{\zeta}
\newcommand{\be}{\begin{equation}}
\newcommand{\ee}{\end{equation}}
\newcommand{\rd}{{\rm d}}
\renewcommand{\k}[1]{{\delta_{#1}}}
\bibliographystyle{prsty}

\title{On random symmetric matrices
with a constraint: the spectral density of random
impedance networks}
\author{J. St\"{a}ring}
\author{B. Mehlig}
\affiliation{School of Physics and Engineering
Physics, Gothenburg University/Chalmers, SE-41296 G\"{o}teborg,
Sweden}
\author{Yan V. Fyodorov}
\affiliation{Department of Mathematical Sciences,
Brunel University, Uxbridge UB83PH, UK}
\affiliation{Petersburg Nuclear Physics Institute,
Gatchina 188350, Russia}
\author{J.M. Luck}
\affiliation{Service de Physique Th\'eorique (URA 2306 of CNRS),
CEA Saclay, 91191 Gif-sur-Yvette cedex, France}
\begin{abstract}
We derive the mean eigenvalue density for symmetric
Gaussian random $N\!\times\!N$ matrices in the limit
of large $N$, with
a constraint implying that the row sum of matrix
elements should vanish. The result is shown to be
equivalent to a result found
recently for the average density of resonances in
random impedance
networks [Y.V. Fyodorov, J. Phys. A: Math. Gen. {\bf
32}, 7429 (1999)].
In the case of banded matrices,
the analytical
results are compared with
those extracted from the numerical solution of
Kirchhoff equations
for quasi one-dimensional random impedance networks.
\end{abstract}

\maketitle

The study of random matrices\cite{meh67} has provided
insight into many physical problems,
both in the quantum and in the classical domain.
For example, random matrices have been very
successfully used
to model statistical properties of disordered
conductors and of highly
excited classically chaotic quantum
systems~\cite{guhr}.
In the classical domain
random matrices arise, for instance,
in the context of diffusion in random, directed
environments (see for example Ref.~\onlinecite{cha97}
and references
therein). In
most of those applications
the random matrix elements obeyed symmetry
requirements where
appropriate, and were otherwise taken to be
independently distributed
random variables.

There are cases, however, where constraints on the
matrix elements must be
considered. Such constraints generate
correlations between the latter:
examples are electron hopping in amorphous
semiconductors
(see Ref.~\onlinecite{mez99} and references cited
therein), random impedance networks~\cite{luck98,fyo99,fyo01}
(see Ref.~\onlinecite{ap} for a review
and Ref.~\onlinecite{cermet} for applications),
and random master equations~\cite{bray88}:
in these cases, the random matrices obey the
constraints
that the row sums of matrix elements should be
zero. This condition
implies correlations between diagonal and off-diagonal
matrix elements.
In Refs.~\onlinecite{mez99,fyo99,bray88}, the average
spectral
density and spectral fluctuations of three different
random-matrix ensembles
of this type were calculated, using the
method of replicas and the supersymmetry approach.

In this paper we derive the
average density of eigenvalues of
a suitably modified ensemble of symmetric Gaussian
random matrices.
The aim is twofold. First, we wish to show that for
full matrices
the density obtained is
equivalent to that found in Refs.~\onlinecite{fyo99}
and~\onlinecite{bray88}. Our second aim is to derive the
corresponding result
for banded matrices, and to compare it with
the results of the numerical solution of the Kirchhoff equations
on random impedance networks with quasi one-dimensional topology.

{\em Formulation of the problem.} We consider an
ensemble
of $N\times N$ random matrices $\bbox{M}$ with matrix
elements
\begin{equation}
\label{eq:defM}
M_{mn} = J_{mn} - \delta_{mn} \sum_{l=1}^N J_{ml},
\end{equation}
where $\bbox{J}$ is a real symmetric $N\times N$
matrix with
random entries. Such ensembles have been considered
in Refs.~\onlinecite{fyo99,fyo01,bray88}, and
\onlinecite{fyo99b}.
In Ref.~\onlinecite{bray88}, $\bbox{M}$ was used to
model
a random transition matrix for a model describing
glassy relaxation, $\partial_t u(t) = -\bbox{M} u(t)$,
with matrix elements $J_{mn}$ distributed
independently (subject to the constraint $J_{mn} =
J_{nm})$
according to
\be
\label{eq:defJ1}
P(J_{mn}) =
\frac{p}{N}\,\delta\!\left(J_{mn}-\frac{1}{p}\right)
+\left(1-\frac{p}{N}\right)\,\delta(J_{mn}).
\ee
The form (\ref{eq:defM}) yields $\sum_{n} M_{mn} = 0$, implying probability
conservation
in the problem considered in Ref.~\onlinecite{bray88}.
The average density $d(\lambda)$ of eigenvalues
$\lambda$,
\be
\label{eq:defd}
d(\lambda) = N^{-1}\langle\mbox{tr}\,
\delta(\bbox{M}-\lambda\bbox{1})\rangle,
\ee
was calculated in the limit of large $N$ and $p$,
using the method of replicas.
$\langle\cdots\rangle$ is an average over the ensemble
defined by $P(J_{mn})$.

In Ref.~\onlinecite{fyo99}, eigenvalues
of matrices ${\bf M}$ with elements
similar to (\ref{eq:defM}) were shown to model
resonance frequencies
in random impedance networks\cite{luck98}, with
$J_{mn} = J_{nm}$ and
\be
P(J_{mn}) = \frac{1}{2}\,\delta\big(J_{mn}-1\big)
          +\frac{1}{2}\,\delta\big(J_{mn}+1\big).
\ee
In Refs.~\onlinecite{fyo99} and \onlinecite{fyo01},
the average density of resonance frequencies was
calculated, in the limit of large $N$,
using a variant of the supersymmetry technique.
It was found that the result agrees~\cite{note} with
that of Ref.~\onlinecite{bray88}, up to
a scale factor (related to $p$) and a rigid shift in
$\lambda$
[related to the fact that $J_{mn} \geq 0$ in
(\ref{eq:defJ1})].

In the following we calculate the ensemble-averaged
density of eigenvalues of ${\bf M}$, treating both
off-diagonal and diagonal
entries of $\bbox{J}$ as independent,
identically distributed Gaussian real variables.
The corresponding
symmetric random matrix ${\bf J}$ belongs to
the Gaussian orthogonal ensemble (GOE)~\cite{meh67}
with joint probability density
\be
\label{eq:defJ3}
P(\bbox{J})\, {\rm d}\bbox{J}\propto
\exp\left(-\frac {1}{4\,\sigma^2} \mbox{tr}\,
\bbox{J}^2\right)\,{\rm d}\bbox{J}.
\ee
The average density of eigenvalues $E$ of such
matrices in the limit
$N\gg1$ is given by the semi-circular law
$d(E) = (2\pi\sigma^2N)^{-1}\sqrt{4\sigma^2 N-E^2}$
for $|E| \leq 2(\sigma^2 N)^{1/2}$ and zero otherwise\cite{footnote}.
In the following we ask 
how imposing a "constraint" (that the row sum of
matrix elements
should be zero) modifies the mean eigenvalue
density of ${\bf M}$
with respect to that of ${\bf J}$.
In the limit of large $N$, the problem may be solved
using diagrammatic perturbation theory (see for
instance
Ref.~\onlinecite{cha97} and references cited therein),
as shown below.

Our results may be summarised as follows. In the
limit of
large $N$, the averaged eigenvalue
density for ${\bf M}$ coincides with the result
derived in Refs.~\onlinecite{fyo99,fyo01}.
In this limit, correlations between diagonal and
off-diagonal
matrix elements are found to be irrelevant, and the
result can be understood
in terms of an averaged Pastur equation\cite{pastur}.
Furthermore, we have also considered the case of
banded symmetric random matrices. This case is of
interest
for random impedance networks with quasi
one-dimensional topology~\cite{luck98,fyo99b}. Our results are 
in good agreement
with
those of exact numerical solutions of the Kirchhoff
equations
for such networks.

\begin{figure}
\centerline{\includegraphics{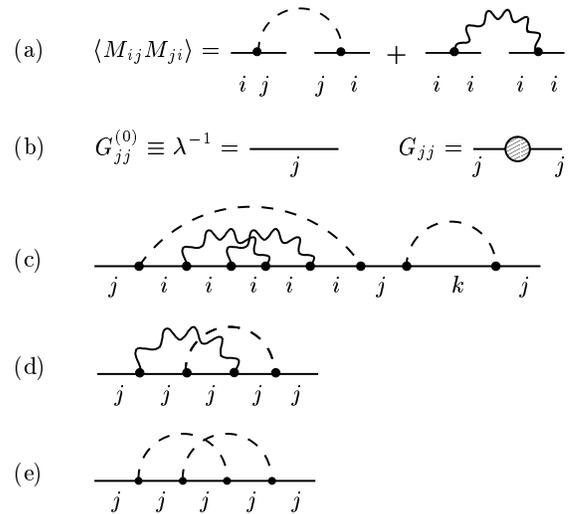}}
\caption{\label{fig:diagrammatics}
(a) Contraction $\langle M_{ij} M_{ji}\rangle$:
the dashed line carries a factor of $1/N$ and
the wavy line carries a factor of unity.
The second term contributes only when $i=j$.
(b) Graphical representation of $G^{(0)}_{jj} \equiv
E^{-1}$
and $G_{jj}$, the $j$th diagonal matrix element of
the average resolvent $\bbox{G}$.
(c) Example of a diagram contributing to $G_{jj}$.
Internal indices are summed over ($i$ and $k$).
(d,e) Diagrams of higher order in $N^{-1}$.}
\end{figure}

{\em Method.} The average eigenvalue density may be
obtained
from the trace of the averaged resolvent
$\bbox{G} =\langle(E\bbox{1}-\bbox{M})^{-1}\rangle$,
\be
d(E) = -(N\pi)^{-1} \mbox{Im}\,
\mbox{tr}\,\bbox{G}.
\ee
Expanding the resolvent,
Wick's theorem may be employed for performing the
average,
using
\begin{eqnarray}
\label{eq:contr1}
&&\langle M_{ij} M_{mn}\rangle
= \sigma^2\Big[
(\k{im}\k{jn} + \k{in}\k{jm})\\
\nonumber
&&\mbox{}\hspace*{0.5cm}-\k{mn}(\k{im}\k{jl}+\k{il}\k{jm})
-\k{ij}(\k{im}\k{kn}+\k{in}\k{km})\\
&&\mbox{}\hspace*{0.5cm}+\k{ij}\k{mn}(N\k{im}+1)\Big].
\nonumber
\end{eqnarray}
In the limit of large $N$, adopting the scaling
$\sigma^2 = N^{-1}$,
(\ref{eq:contr1}) can be simplified to
\be
\label{eq:contr2}
\langle M_{ij} M_{mn}\rangle
\simeq \frac{1}{N} (\k{im}\k{jn} + \k{in}\k{jm}) +
\k{ij}\k{mn}\k{im}.
\ee
It is convenient to keep track of the contributions
with the help of diagrams. To this end, a graphical
representation
of the contraction (\ref{eq:contr2}) is introduced
in Fig.~\ref{fig:diagrammatics}(a).
The averaged resolvent $\bbox{G}$
turns out to be a diagonal matrix.
Fig.~\ref{fig:diagrammatics}(b) defines
a graphical representation for the diagonal elements
$G_{jj}$
of $\bbox{G}$. Terms contributing to $G_{jj}$
are shown in Fig.~\ref{fig:diagrammatics}(c-e).
One observes that, in the limit of large $N$, only
diagrams with no intersections between dashed
or between dashed and wavy lines contribute.
Thus, to leading order in $N$, the contribution
(c) in Fig.~\ref{fig:diagrammatics} must be
considered,
but not (d) or (e), for example.

\begin{figure}
\centerline{\includegraphics{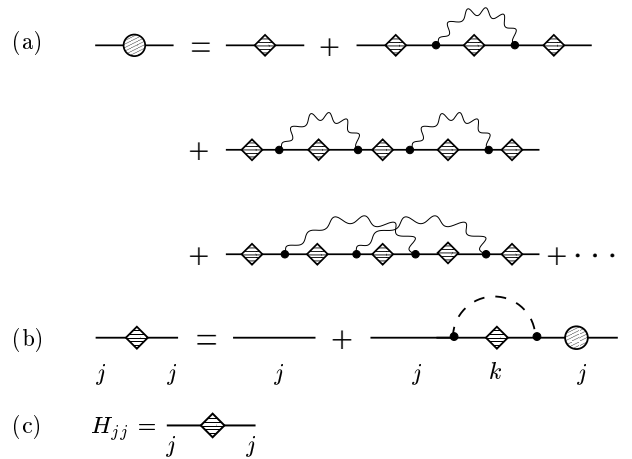}}
\caption{\label{fig:dyson} (a,b) Self-consistent
equations for the average of the resolvent. (c)
Graphical representation of the auxiliary functions
$H_{jj}$.}
\end{figure}

All diagrams contributing to the resolvent to leading
order in $N^{-1}$ may be summed, resulting in a set
of two coupled equations shown in
Fig.~\ref{fig:dyson}(a) and (b). The diagonal elements
$G_{jj}$ and
$H_{jj}$ are independent of $j$ and denoted by $G$ and
$H$
in the following.
The first equation [Fig.~\ref{fig:dyson}(a)] contains
an infinite
sum of diagrams. This sum may be performed exactly.
It is denoted by $g(H^{-1})$, where
\begin{equation}
g(z) = (2\pi)^{-1/2} \int_{-\infty}^{\infty}\!\rd
J\,\frac{\exp(-J^2/2)}{z-J}.
\end{equation}
The result for the resolvent scales with $\sigma^2$ and $N$ as
$(\sigma^2 N)^{-1/2}\,g(H^{-1}\,(\sigma^2N)^{1/2})$.
Consequently,
the equations shown in Fig.~\ref{fig:dyson}(a,b) imply
the following
self-con\-sis\-tent equation
\be
\label{eq:sc}
G = (\sigma^2 N)^{-1/2}\,g\big((E-G)\,(\sigma^2
N)^{1/2}\big),
\ee
equivalent to Eq.~(51) in Ref.~\onlinecite{fyo99} for
the
average density of resonances in an ensemble of
highly connected random impedance
networks\cite{note1}.

\begin{figure}
\psfrag{x}{$E/\sqrt{b}$}
\psfrag{y}{$d(E)\,\sqrt{b}$}
\centerline{\includegraphics[width=6.cm,clip]{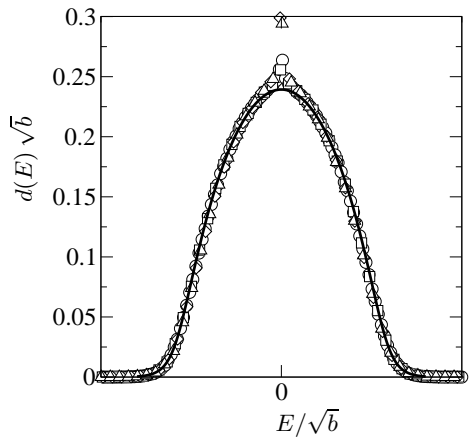}}
\caption{\label{fig:rmt} Density of eigenvalues of
exact diagonalisations of random matrices of the form
(1,5) for $\sigma=1$, $N=500,1000$, and $b=50,100$ (symbols),
together with the prediction (\ref{eq:sc2}) (solid line).}
\end{figure}

The form of the simplified
contraction (\ref{eq:contr2}) implies
an interpretation of the result (\ref{eq:sc}) in
terms
of an averaged Pastur equation: consider a random
matrix $\mathbf{M} =
\mathbf{J}+\mathbf{V}$, where $\mathbf{J}$ is
distributed according to
(\ref{eq:defJ3}), and $\mathbf{V}$ is a diagonal
matrix with
Gaussian random entries $v_k$ with zero mean and unit
variance,
independent of $J_{mn}$. For a given realisation of
$\mathbf{V}$, Pastur's equation\cite{pastur} is
$\mathbf{G} = (E\mathbf{1}
-\mathbf{V}-\mathbf{G})^{-1}$
(see also Ref.~\onlinecite{fyo95} and references
therein). In this equation $\mathbf{G}$
is the $\mathbf{J}$-averaged resolvent, keeping
$\mathbf{V}$ fixed.
One obtains Eq.~(\ref{eq:sc}) after observing that
$\mathbf{G}$ is diagonal,
by averaging over the matrix elements of $\mathbf{V}$.
This interpretation implies that, in the limit of large
$N$,
the correlations between diagonal and off-diagonal
matrix elements
of $\mathbf{M}$ [as seen in
Eqs.~(\ref{eq:defM},\ref{eq:contr1})]
are irrelevant.

The above procedure is easily extended to the case
of banded matrices, also of interest in random
im\-pe\-dance networks~\cite{luck98,fyo99b}. In the banded case, $\langle
J_{mn} J_{kl} \rangle =
\sigma^2 (\delta_{mk}
\delta_{nl}+\delta_{ml}\delta_{nk})$
[which follows from (\ref{eq:defJ3})] is replaced by
\be
\langle J_{mn} J_{kl} \rangle =
\sigma^2(|m-n|)\,\, \left(\delta_{mk}
\delta_{nl}+\delta_{ml}\delta_{nk}
\right).
\ee
The function $\sigma^2(x)$ is given by
\be
\sigma^2(x) = \left\{\begin{array}{ll}
\sigma^2 & \mbox{for $0 \leq x \leq b/2$},\\
0 & \mbox{otherwise}.
\end{array}
\right.
\ee
The bandwidth of $\bbox{J}$ is thus $b$.
In the limit of large $N$ and large $b$,
the spectral density is given by a slight
modification of (\ref{eq:sc}),
\be
\label{eq:sc2}
G = (\sigma^2 b)^{-1/2}\, g\big((E-G)\, (\sigma^2
b)^{1/2}\big).
\ee
Diagrammatically,
the necessary changes are most easily
derived by letting $\sigma^2 = N^{-1}$ and
assuming that $b = B N$ (with fixed $B\ll 1$).
Then the wavy line in Fig.~\ref{fig:diagrammatics}(a)
acquires a factor of $B$.
Furthermore, the dashed line in Eq.~(\ref{eq:contr2})
also acquires a factor of $B$.
Hence the self-consistency
equation in the banded case becomes (\ref{eq:sc2}).
Let us also note that the same formula is actually
valid not only for $b\sim N$, but more generally for $1\ll
b\ll N$.

This result implies that densities for
different values of $b$ can be scaled
on one single curve by plotting $\sqrt{b}\,
d(E/\sqrt{b})$.
In Fig.~\ref{fig:rmt}, solutions of (\ref{eq:sc2}) for $\sigma=1$
are compared with results of exact diagonalisations
of random matrices with $N=500,1000$ and $b=50,100$.
We observe a very good agreement.
The results confirm that, for large $N$ and $b$,
the average density of eigenvalues is independent of $N$, and that it scales with $b$
as expected.

In the following we show that (\ref{eq:sc2}) also
describes the density of resonances
for certain random impedance networks with a large, but finite connectivity.

\begin{figure}
\psfrag{TAG4}{$n-b/2$}
\psfrag{TAG3}{$n-1$}
\psfrag{TAG2}{$n$}
\psfrag{TAG1}{$n\!+\!1$}
\psfrag{TAGN}{$n\!+\!b/2$}
\mbox{}\vspace*{1cm}
\centerline{\includegraphics[width=6.cm,clip]{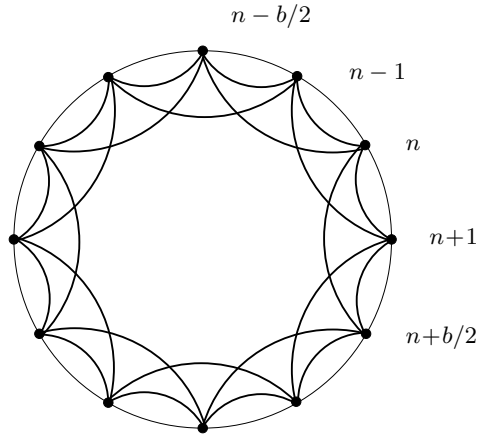}}
\caption{\label{fig:lattice} Topology of a (quasi)
one-dimensional lattice
with periodic boundary conditions. Each site $n$ is
connected to its $b=4$ neighbours.}
\end{figure}

\begin{figure}[t]
\psfrag{x}{$E/\sqrt{b}$}
\psfrag{y}{$d(E)\,\sqrt{b}$}
\centerline{\includegraphics[width=6.cm,clip]{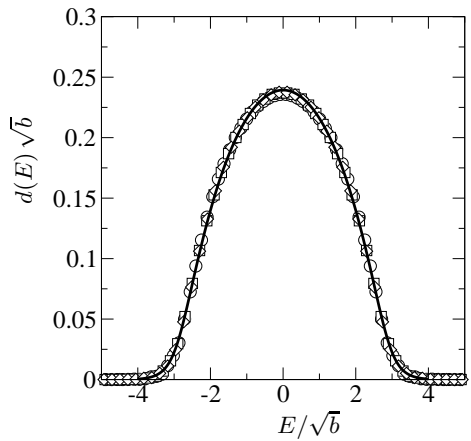}}
\caption{\label{fig2}
Plot of the density of resonances for a
quasi one-dimensional random impedance network
(see text).
Data for three different ranges
($b=60$, $100$, and $120$) (symbols) are compared
with the theoretical prediction (\ref{eq:sc2}).}
\end{figure}

{\em Random impedance networks.} Random networks of complex impedances are currently
used to model
electrical and optical properties of disordered
inhomogeneous media~\cite{ap}.
The most common situation is that of a binary
composite medium,
modeled by attributing a random conductance to each
bond $({\bf x},{\bf y})$ of a lattice, according to the binary law:
\begin{equation}
\sigma_{{\bf x},{\bf y}}=\left\{\matrix{
\sigma_0&\hbox{with probability}&p,\hfill\cr
\sigma_1&\hbox{with probability}&q=1-p.\hfill\cr
}\right.
\end{equation}
The homogeneous Kirchhoff equations for the electric potentials,
\begin{equation}
\sum_{\bf y}\sigma_{{\bf x},{\bf y}}(V_{\bf y}-V_{\bf
x})=0,
\end{equation}
can be recast as
\begin{equation}
(\Delta_Q-\lambda\Delta)V=0,
\label{homo}
\end{equation}
with $\lambda={\sigma_0}/({\sigma_0-\sigma_1})$, and
\begin{eqnarray}
\nonumber
(\Delta V)_{\bf x}&=&\sum_{{\bf y}({\bf x})}(V_{\bf
y}-V_{\bf x}),\\
(\Delta_P V)_{\bf x}&=&\sum_{{\bf y}\in P({\bf
x})}(V_{\bf y}-V_{\bf
x}),\\
(\Delta_Q V)_{\bf x}&=&\sum_{{\bf y}\in Q({\bf
x})}(V_{\bf y}-V_{\bf x}),
\nonumber
\end{eqnarray}
where ${\bf y}({\bf x})$ are all the sites connected
to site ${\bf x}$,
whereas ${\bf y}\in P({\bf x})$ (resp. ${\bf y}\in Q({\bf x})$)
are those connected by a conductance $\sigma_0$ (resp. $\sigma_1$),
so that $\Delta=\Delta_P+\Delta_Q$.
Resonances appear as non-trivial solutions to Eq.~(\ref{homo}),
for $0<\lambda<1$.

An efficient algorithm
allowing for an exact determination of all the
resonances
of a finite sample
has been developed in Ref.~[5].
We have adapted this algorithm to the simplest geometry
allowing for long-ranged bonds.
Each site $n$ of a very long chain is connected to its $b$
neighbors $(n-b/2,\dots,n-1,n+1,\dots,n+b/2)$, 
as shown in Fig.~\ref{fig:lattice}.
Periodic boundary conditions are assumed.
For definiteness we choose $p=q=1/2$,
so that all the resonances are expected to be located at 
$\lambda=1/2$ in the $b\to\infty$ limit.

Our numerical results are shown in Fig.~\ref{fig2},
for very long periodic chains with ranges 
$b=60$, $100$, and $120$.
For each value of $b$, we have accumulated a number of resonances of order
$10^7$.
After rescaling the resonances according to
$
\lambda = \left(1+E\right)/2$
the density is given by (\ref{eq:sc2}) with $\sigma=1$.
A very satisfactory quantitative agreement with the
theoretical prediction is observed.

{\em Acknowledgements.} B.M. gratefully acknowledges
discussions with J.T. Chalker. This work was in part
supported by Vetenskapsr\aa{}det (J.S. and B.M.), 
and by EPSRC Research Grant No. GR/13838/01
(Y.V.F).


\begin{thebibliography}{10}
\bibitem{note}
The average density of resonance frequencies of an
ensemble
of random impedance networks is obtained from a
generalised
eigenvalue problem (random matrix pencil) of the form
$\bbox{M} u = \lambda \bbox{W} u$,
with $W_{mn} = (N+2)\delta_{mn}-1$.
It was shown~\protect\cite{fyo99,fyo01} that,
as a consequence of the particular
structure [see eq.~(\ref{eq:defM})] of $\bbox{M}$, and of the
form of $\bbox{W}$,
the latter can be replaced by $N$ times the unit
matrix without
changing the average density of eigenvalues.
\bibitem{footnote}
The constant of proportionality is fixed by normalisation.
To compare with Refs.~\onlinecite{luck98} and \onlinecite{ap},
the variance should be set unity; for the perturbative
calculations described below it is convenient to take $\sigma^2 = N^{-1}$.
\end{thebibliography}
\end{document}